\documentclass[aps,rmp,reprint,superscriptaddress,floatfix]{revtex4-1}
\usepackage[utf8]{inputenc}
\usepackage[T1]{fontenc}
\usepackage[usenames,dvipsnames,svgnames,table]{xcolor}
\usepackage{lmodern}
\usepackage{graphicx}
\usepackage{amsmath}
\usepackage{hyperref}
\definecolor{darkblue}{rgb}{0.0,0.0,0.3}
\hypersetup{colorlinks=true,citecolor=gray,urlcolor=darkblue}

\makeatletter
\let\cat@comma@active\@empty 

\renewcommand\@make@capt@title[2]{%
 \@ifx@empty\float@link{\@firstofone}{\expandafter\href\expandafter{\float@link}}%
  {\textbf{#1}}\@caption@fignum@sep#2\quad
}%

\renewcommand{\fnum@figure}{\textbf{Figure~\thefigure}.}
\makeatother

\setcitestyle{numbers,square,comma}
\begin{document}


\title{Synthesis and characterization of attosecond light vortices in the extreme ultraviolet}

\author{R. Géneaux}
\affiliation{LIDYL, CEA, CNRS, Université Paris-Saclay, CEA Saclay, 91191 Gif-sur-Yvette, France.}
\author{A. Camper}
\affiliation{Department of Physics, Ohio State University, Columbus, Ohio 43210, USA.}
\author{T. Auguste}
\affiliation{LIDYL, CEA, CNRS, Université Paris-Saclay, CEA Saclay, 91191 Gif-sur-Yvette, France.}
\author{O. Gobert}
\affiliation{LIDYL, CEA, CNRS, Université Paris-Saclay, CEA Saclay, 91191 Gif-sur-Yvette, France.}
\author{J. Caillat}
\affiliation{ Sorbonne Universités, Laboratoire de Chimie Physique—Matière et Rayonnement (UMR7614), UPMC Univ Paris 06, 11 rue Pierre et Marie Curie, 75005 Paris, France.}
\author{R. Taeïb}
\affiliation{ Sorbonne Universités, Laboratoire de Chimie Physique—Matière et Rayonnement (UMR7614), UPMC Univ Paris 06, 11 rue Pierre et Marie Curie, 75005 Paris, France.}
\author{T. Ruchon}
\email[Corresponding Author: ]{thierry.ruchon@cea.fr}
\affiliation{LIDYL, CEA, CNRS, Université Paris-Saclay, CEA Saclay, 91191 Gif-sur-Yvette, France.}

\date{\today}

\begin{abstract}
Infrared and visible (IR-Vis) light beams carrying orbital angular momentum (OAM) are currently thoroughly studied for their extremely broad applicative prospects, among which quantum information, micro-machining and diagnostic tools. Here, we largely extend these prospects, presenting a comprehensive study for the synthesis and full characterization of optical vortices carrying OAM in the extreme ultraviolet (XUV) domain. We establish the upconversion rules of a femtosecond IR-Vis helically phased beam into its high order harmonics, showing that each harmonic order carries the total number of OAM units absorbed in the process. To demonstrate a typical use of these new XUV light beams, we show our ability to generate, through photoionization, and control attosecond electronic wavepackets carrying OAM. These unique attosecond light and electron springs are promising secondary sources for applications and fundamental tests.
\end{abstract}


\maketitle

\section{Introduction}
Like any massive particle may carry two types of angular momenta (AM), namely spin and orbital angular momenta (SAM and OAM respectively), the massless photon also possesses two types of angular momentum \cite{landau1982a}. It was early recognized that SAM is associated to the circular polarization of light beams but, only lately was the OAM of light beams associated with tilted wave front beams and made largely accessible experimentally in the infrared and visible (IR-Vis) spectral regions \cite{allenpra1992}. Helically phased light beams with a phase singularity associated to zero intensity along the axis are the most common implementations of macroscopic beams carrying OAM \cite{yaoaop2011}. At the photon level, the twisted wavefront of the beam translates into a quantized momentum, with its component along the beam axis taking discrete  $\ell\hbar$ values, where $\ell$ is any positive or negative integer number \cite{yaoaop2011}. The unique properties of twisted light beams which are showing up both at the macroscopic and photon levels, lead to the emergence of countless applications, ranging in the Vis-IR domain from quantum information \cite{wangnp2012} to microscopy \cite{furhapterol2005}, nanoparticles manipulation \cite{yaoaop2011} or fine structuring of materials using pulsed lasers \cite{toyodananolett2012}. In the XUV domain, twisted light beams hold the potential for specific fundamental and applied prospects, for instance new kind of dichroisms \cite{vanveenendaalprl2007,zambrananc2014} or the visualisation of dislocation strain fields in bulk material \cite{takahashiprb2013}. These new perspectives have already motivated the development of XUV sources carrying OAM on both quasi-continuous synchrotron installations \cite{bahrdtprl2013}, and on femtosecond free electron lasers \cite{hemsingnp2013,ribiprl2014}. One of the main outcomes reported here is the observation of such beams both in the \textit{attosecond} regime (1 as = $10^{-18}$ s) and in the extreme ultraviolet (XUV) domain.

As a table top, ultra-short and largely tunable alternative to large scale instruments, High-Harmonic Generation (HHG) based XUV sources show unrivalled stability specifications, especially useful for pump-probe experiments targeting attosecond time resolution \cite{corkum2007}. They are based on the upconversion of a high intensity femtosecond Vis-IR laser beam into the XUV range through nonlinear interaction with a gas. Very recently, two groups reported HHG with helically phased light beams \cite{zurchnp2012,gariepyprl2014}. However, the conclusions of these two papers as for the harmonics’ OAM are in strong disagreement. On the one hand, analysing a couple of low HHG orders, Gariepy et al. \cite{gariepyprl2014} confirmed the theory of Hernández-García el al. \cite{hernandezprl2013} and the general “multiplicative” rule for OAM transfer in nonlinear processes \cite{yaoaop2011}: for the two harmonic orders they could characterize, they observed that the OAM of harmonic order q is the quantum number $\ell_q=q\ell_1$, where $\ell_1$ is the OAM of the driving field. On the other hand, Zürch et al. \cite{zurchnp2012} observed on a single harmonic that $\ell_q=\ell_1$, arguing that parametric instabilities prevent higher order vortices from propagating towards the detector. XUV light pulses of attosecond duration carrying OAM, based on a broad comb of phase-locked helically-phased harmonics, thus remain to be observed experimentally. Such \textit{attosecond} light springs \cite{parienteol2015}, lying in the XUV spectral range, would be a unique source to tailor \textit{attosecond} electron beams carrying OAM through photoionization. Twisted electron beams are actively studied for future applications in fields covering spectroscopy of diluted and condensed-matter \cite{asenjoprl2014}, microscopy and particle physics \cite{grilloapl2014}.  Currently, they are primarily generated through tailoring non-singular electron beams in the \textit{quasi-static} regime \cite{ushidan2010, verbeeckn2010}. \\

In the present work, we first report on the validity of the multiplicative OAM transfer rule for HHG with infrared helically phased beams over an extremely broad spectrum and in several conditions of practical interest. Thus, we establish the opportunity to synthesise XUV “light springs” with attosecond structures. We additionally use these light springs to generate attosecond electron beams with vortices, which were fully characterized in time. Finally, we propose some promising applications for these beams such as tests of the photoionization selection rules as recently theoretically discussed \cite{piconoe2010, matulajpb2013}. 

\section{Multiplicative rule for OAM transfer through HHG}
As raised up by Gariepy et al. \cite{gariepyprl2014}, the main difficulty encountered when studying XUV twisted attosecond pulses lies in the need for OAM characterization over a broad spectral range. In the Vis-IR range, the phase fronts of twisted light beams are usually analysed directly using diffractive optical elements or interference schemes to determine the amount of OAM \cite{yaoaop2011}. Zürch et al. \cite{zurchnp2012} and Gariepy et al. \cite{gariepyprl2014} transferred these approaches to the XUV domain, but soon faced the challenge of having to resolve fringes with extremely low spacing due to the short wavelength of the XUV. In practice, the OAM of only one or two low HHG orders were analysed. However, these light beams show ring-shaped intensity profiles while propagating, which is another signature of their OAM. Here, we exploit the parameters of these peculiar profiles for their OAM measurements. \\

To illustrate our approach, we focus on Laguerre-Gaussian (LG) modes, which are eigen solutions of the paraxial wave equation and form a natural set of helically phased light beams carrying all possible integer values of OAM \cite{yaoaop2011}.  We restrict ourselves to LG modes with a radial index $p=0$, thus showing a single intensity ring. With the simple hypothesis that HHG driven by such an IR LG beam with OAM $\ell_1$ leads to a comb of XUV LG modes, with the q-th order carrying an OAM, the radius of the XUV beam reads \cite{suppmat}
\begin{equation}
\label{eq:one}
{{r}_{max}}\left( {{\lambda }_{\text{q}}},z \right)={{w}_{0}}\sqrt{\frac{\left| {{\ell }_{1}} \right|}{2}}.\sqrt{1+{{\left( \frac{\pi z\text{ }\!\!~\!\!\text{ }{{w}_{0}}^{2}}{{{\lambda }_{1}}}\frac{q{{\ell }_{1}}}{{{\ell }_{q}}} \right)}^{2}}},
\end{equation}
where $z$ is the distance along the optical axis from the focal point and $w_0$ is the equivalent of the waist for a Gaussian beam. Eq.~(\ref{eq:one}) establishes that the value of $\ell_q$ rules the order dependence of the harmonics rings radii. In particular, the multiplicative law ${{\ell }_{q}}=q\times {{\ell }_{1}}$ yields a constant ring diameter over the whole spectrum while ${{\ell }_{q}}={{\ell }_{1}}$ $\left( {{\ell }_{1}}\ne 0 \right)$ would lead to a diameter increasing with q. It is worth noting that none of these predictions matches the behavior observed when driving HHG with regular Gaussian beams, which leads, for the central part of the XUV beam, to an increasing divergence with q. This regular behavior is mainly a consequence of the HHG process itself combined to the $\lambda$-dependence of light propagation to the far field \cite{suppmat}. Here we predict that the helical phase would modify this regular behavior. Inspecting the divergence of harmonics should thus answer the controversy about the transfer law, without resorting to any diffractive element or interference scheme. \\

\begin{figure*}
		\includegraphics[scale=0.8]{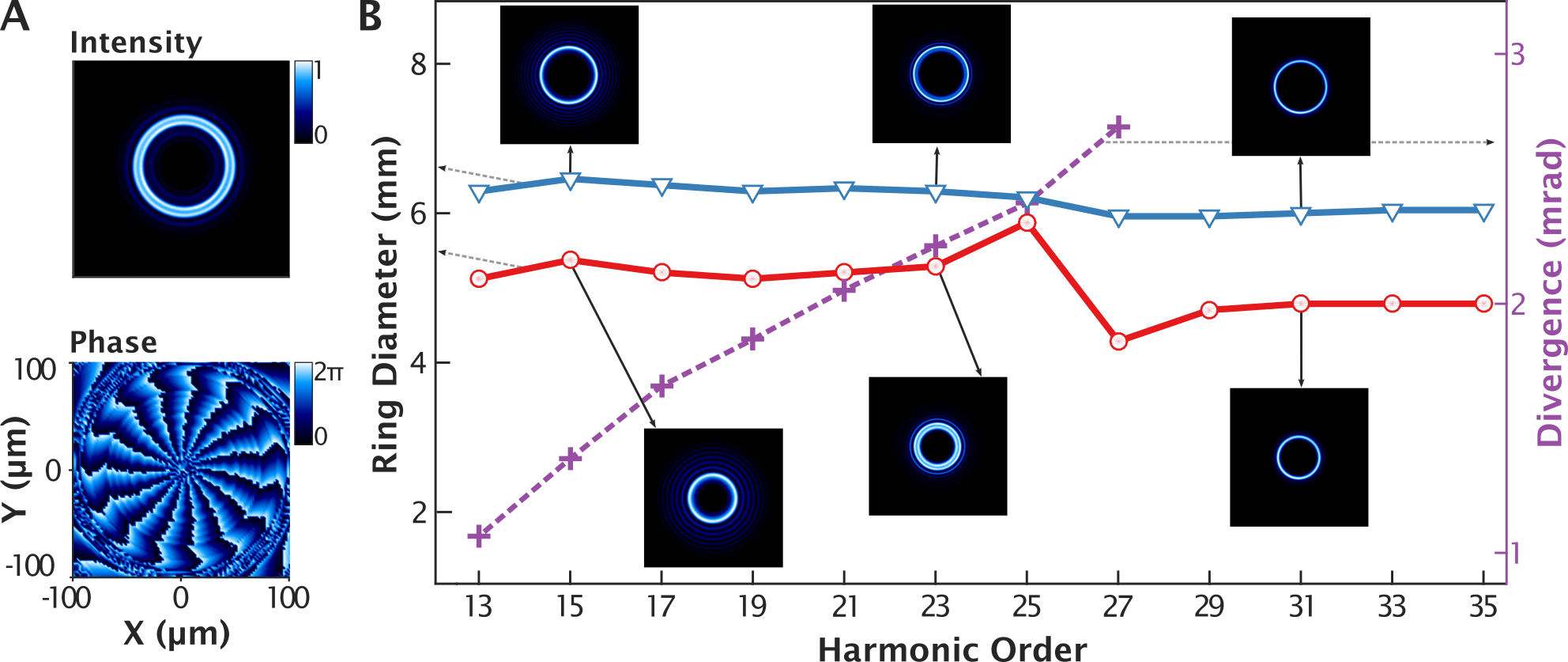}
		\caption{\label{fig:one}\textbf{Full calculation of high order harmonic profiles generated with a driving laser carrying OAM.}\\ \textbf{(A)} Intensity and phase transverse profiles of harmonic 15 at the exit of the medium. We consider a spatially and temporally incident Gaussian beam of 6.25 mm waist and 50 fs FWHM duration, focused by a 1 m focal length lens, in the middle of a 500 $\mathrm{\mu m}$ wide (FWHM) Lorentzian argon jet. The equivalent Gaussian beam waist at focus $w_0$ is 40 $\mathrm{\mu m}$, leading to an equivalent Rayleigh range $z_R$ of 6.5 mm. The maximum gas pressure is 10 mbar and the laser peak intensity at focus is $\mathrm{1.5\times 10^{14} W/cm^2}$. The observation plane is located 80 cm downstream of the jet exit. \textbf{(B)} Diameter of the harmonic rings in the far field for $\ell_1=1$ (red circles) and $\ell_1=1$ (blue triangles). The data for $\ell_1=0$ (purple crosses) is the short trajectory divergence experimentally measured using a Gaussian beam and a 2 m focal length lens. Insets: corresponding intensity profiles for a selection of harmonics.}
\end{figure*}

The above considerations rely on the hypothesis that a pure LG mode is generated through HHG. To provide more steady grounds to this assumption, we performed a numerical simulation of HHG in argon gas within the strong field approximation (SFA), including macroscopic propagation, in a similar way as in \cite{camperpra2014} (see \cite{suppmat}). The results of the calculations are reported in Figure~\ref{fig:one}. As expected, HHG occurs along the high intensity ring of the generating IR beam. The full width at half maximum (FWHM) of the profile obtained by a cut along a radius of this ring (hereafter called the thickness of the ring) is a fraction of the thickness of the ring of the driving field, a consequence of the high non-linearity of HHG. As for the spatial phase, a typical example is given by the 15$^{th}$ harmonic order (H15), showing a spiral running 15 times $2\pi$ along the ring. This implies that the helicity of H15 is 15 times the helicity of the driving field and that H15 photons carry 15 times the OAM of the fundamental frequency photons. This behavior is consistent with the multiplicative law of nonlinear optics, and was observed here for all computed harmonics, from H11 to H33. To mimic the experiment, we simulate the propagation of the XUV beam towards an observation plane located 80 cm downstream the gas cell. When reaching the detector the harmonic beams still display a ring shape, suggesting that generation at focus is dominated by the emission of a \textit{single} LG mode. As predicted by Eq.~(\ref{eq:one}) with ${{\ell }_{q}}=q\times {{\ell }_{1}}$, the ring diameter is identical for all harmonic orders. It only depends on the OAM value of the driving field, which affects the diameter of the ring. The latter increases by a factor of $1.4\simeq \sqrt{(2)}$ when doubling the helicity of the phase, consistently with the facts that the OAM carried by the harmonics is doubled and that the divergence goes like $\sqrt{\left| {{\ell }_{1}} \right|}$, according to Eq.~(\ref{eq:one}).\\

Getting back to the simplest microscopic model of HHG, by which an electron tunnels out of an atom close to an extremum of the driving field, then acquires energy by quivering in this field and finally recombines with its parent ion emitting its excess of energy as XUV radiation \cite{corkum2007}, it may be anticipated that HHG does not generate pure LG beams, as pointed out by Zürch et al. (14). Indeed, in the full computation, extra faint rings appear on the outermost part of the beams. They are decreasingly divergent with harmonic order until they merge with the central one \cite{suppmat}. Such patterns were lately analyzed for HHG with regular Gaussian beams (${{\ell }_{1}}=0$) and attributed, depending on the generation conditions, to either diffractive effects, Maker fringes in the time domain \cite{heylprl2011} or interferences of the emission from different quantum paths during the excursion of the electron around the ionic core \cite{zairprl2008}.  By numerically turning off these trajectories one at a time, we observed that the outer rings are mostly present in the contribution of the more divergent long one. Being in a loose focusing configuration, this allows us to attribute them mostly to diffraction effects, rather than interferences between the two quantum paths, thus letting the main conclusion of this paragraph unchanged.

\section{Attosecond XUV light springs}
\begin{figure}[ht]
		\includegraphics[width=0.48\textwidth]{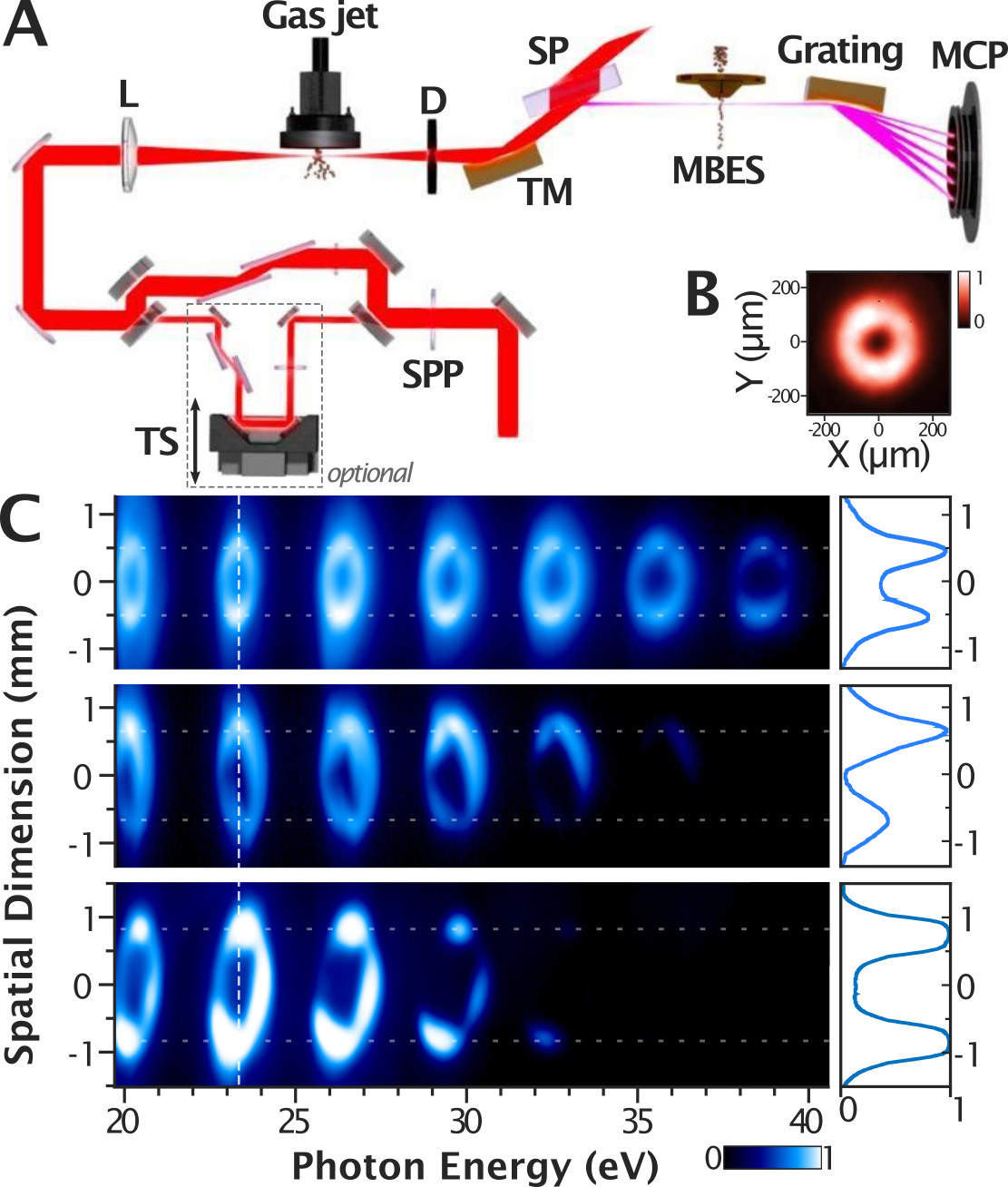}
		\caption{\label{fig:two}\textbf{Experimental observation of HHG spectra carrying OAM.} \textbf{(A)} Experimental setup. It has two operation modes, corresponding to the imaging of HHG light spectra and to the characterization of electronic wavepackets. The Ti:Sapphire laser beam is passed through a spiral phase plate (SPP) and used to generate harmonics. They can be either directly imaged using a photon spectrometer (Grating+Micro Channel Plate) or focused inside a magnetic bottle electron spectrometer time of flight (MBES). An extra weak IR beam may be superimposed with the XUV beam in the MBES with a delay controlled by a translation stage (TS), allowing us to perform RABBIT measurements. See \cite{suppmat} for more details. \textbf{(B)}Intensity profile of the laser beam at focus close to the HHG gas inlet. \textbf{(C)} Normalized intensity of harmonics $\mathrm{15th}$ to $\mathrm{27th}$ generated in argon and observed in the far field, using $\ell_1$ = 1 (top row), $\ell_1$  = 2 (middle row) and $\ell_1$  = 3 (bottom row). }
\end{figure}

The numerical results above were tested experimentally on the LUCA laser server in Saclay. The experimental setup is described in Figure~\ref{fig:two} (A). The laser beam is spatially filtered to yield a very close-to-Gaussian input mode \cite{mahieuapb2015}. A spiral phase mask is inserted to impose a staircase phase profile on this incoming beam, which can either be 1, 2 or 3 times $\mathrm{2\pi}$ per turn, using two successive masks. Focusing such a phase shaped beam leads to a distribution of light about the focal point very close to a LG mode carrying respectively $\ell_1=\pm1,\pm2,\pm3$ units of OAM, the sign being determined by the orientation of the phase mask \cite{beijersbergenoc1994}. 
This is illustrated by the intensity profile reported in Figure~\ref{fig:two} (B), the characterization of the OAM using a diffraction technique being described in \cite{suppmat}. Harmonic spectra obtained using an IR driving field with $\ell_1=$1, 2, 3 are displayed in Figure~\ref{fig:two} (C). For each harmonic order and all OAM values of $\ell_1$, we observe a clear ring shape pinched in the horizontal dimension by the dispersion of the grating. The divergence of the harmonics, seen in the vertical dimension, appears to be constant throughout the whole spectrum, in contrast to what is commonly observed with a Gaussian beam. We measured the average ring diameters to be $1.01\pm 0.02$ mm, $1.33\pm 0.04$ mm and $1.61\pm 0.01$ mm for $\ell_1 = $ 1, 2 and 3, respectively. This is in agreement within 5\% with the $\sqrt{\left| {{\ell }_{1}} \right|}$ dependency predicted by Eq.~(\ref{eq:one}), which leads to ratios of  $\sqrt{2}\simeq 1.4$ and $\sqrt{3}\simeq 1.7$ between the diameters. Note that among all transfer laws we tested; only the multiplicative law provides such a quantitative agreement with the observations \cite{suppmat}. This extends the conclusions of Gariepy et al. (15) to the \textit{whole harmonic spectrum}, opening the way to the synthesis of attosecond light pulses carrying extremely high values of OAM.\\

Furthermore, it appears in Figure~\ref{fig:two} (C) that the highest observed harmonic order decreases with increasing $\ell_1$. This drop in the cutoff energy is due to a lower IR peak intensity at focus, that can be attributed first to a lower efficiency in the conversion from the IR Gaussian mode to the IR LG mode \cite{beijersbergenoc1994}. Second, the IR focal spot size increases with $\ell_1$, requiring a larger aperture of the IR beam to reach the nominal intensity for HHG. This is very demanding on the stability of the incoming phase profile of the IR field and, in practice, leads to imperfect intensity and phase profiles at focus. This translates into a lower yield of the XUV radiation and a degradation of the ring shape when increasing $\ell_1$. However, the highest value of $\ell_q$ is obtained for $\ell_1$=3 with harmonic 19, yielding $\ell_q$=57. Moreover, to get fairly high $\ell_q$ while keeping a smooth intensity profile, we observed that it is beneficial to rather use neon gas as a target with  $\ell_1$=1. The high ionization potential of this gas allows to couple a lot of energy into the target before its ionization is saturated while, due to the low $\ell_1$, it remains less sensitive to wavefront imperfections. As shown in \cite{suppmat}, in these conditions we could generate harmonics up to the $\mathrm{41^{st}}$ order with a smooth intensity profile. Once again, we observed a constant divergence of the spectrum, confirming the general validity of our first measurements in argon. 

\section{Attosecond electron springs}
\begin{figure*}
		\includegraphics[width=0.8\textwidth]{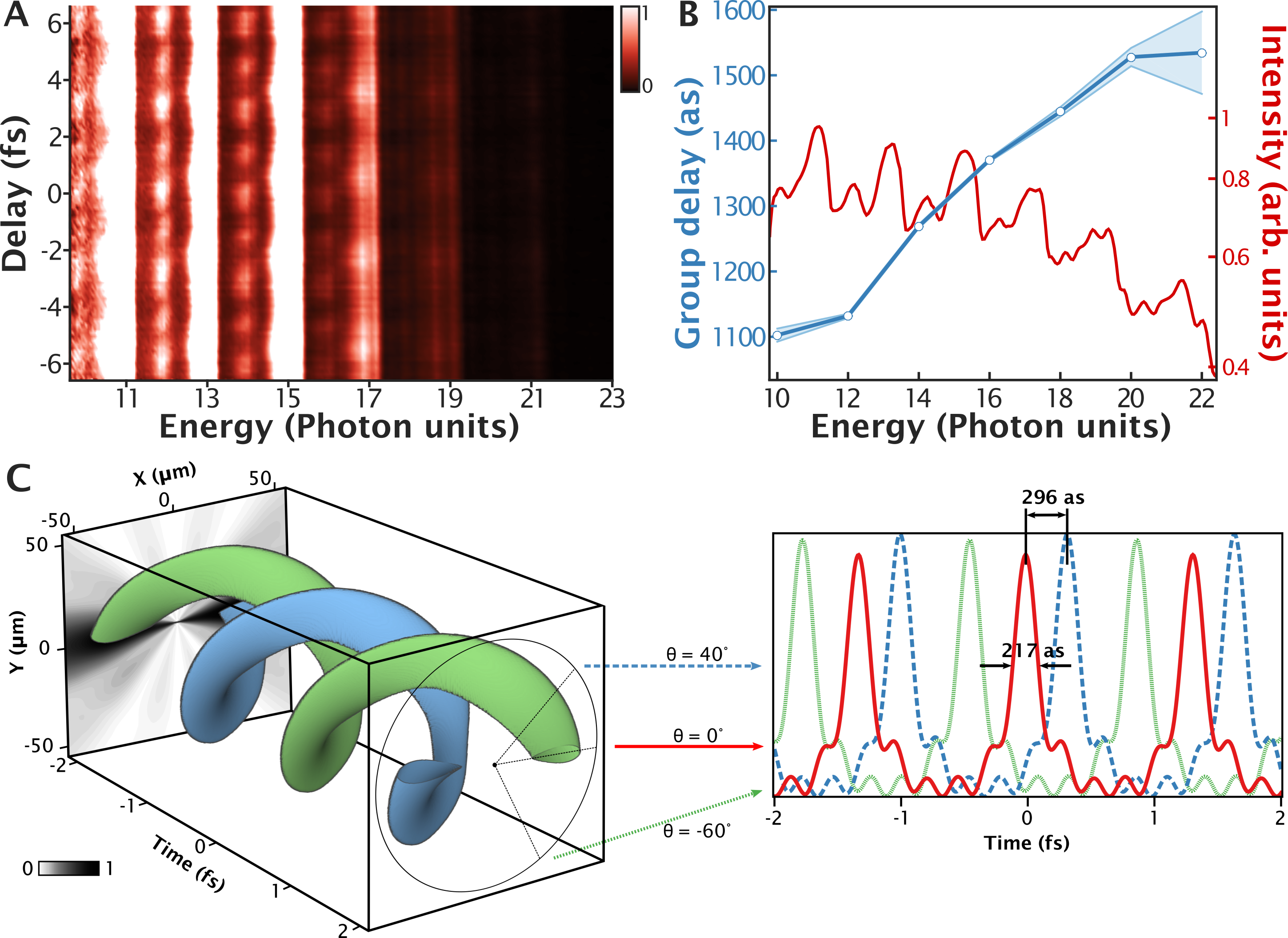}
		\caption{\label{fig:three}\textbf{Attosecond electronic beams carrying OAM. } \textbf{(A)} Two-color XUV+IR two-photon photoionization spectrogram of argon for a driving field with ${\ell _1}$=1. The main lines correspond to odd harmonics while the weaker oscillating ones, showing a double periodicity of T=1.33 fs and 2.7 fs are sidebands. The dynamical range of our spectrometer limits the number of measurable sidebands even if the harmonic cut-off lies higher. \textbf{(B)} Delay-averaged spectrum (red) in log scale and group delay of the electronic wave packets emitted (blue circles, the light blue strip about the curve represents the numerical analysis error bar at 3$\sigma$). The analysis of the sidebands is carried out with the theory of RABBIT \cite{mairessescience2003}, under the assumption of fully coherent light (34). \textbf{(C)} Spatio-temporal shape of the emitted electron beam using the intensity profiles from Fig. 2 and the group delays from panel (B). The 3D surface is a contour at 80\% of the max intensity. The back panel is a projection of this intensity. (inset) Temporal cuts at three different azimuthal angles in the electron beam $\theta=0^{\circ}$(red), $40^{\circ}$(blue) and $-60^{\circ}$(green).}
\end{figure*}

Since the early days of attophysics it has been recognized that XUV attosecond pulses could tailor electron wave packets (EWP) through photoionization \cite{corkum2007,caillatprl2011}. Applications of such attosecond EWP were proposed, for instance, as a quantum stroboscope \cite{mauritssonprl2006}, or as tools to localize EWP in space and time around atoms and molecules \cite{calegari2014}. In the present context, we propose to increase the tailoring knobs by transferring the OAM carried by harmonics onto photoionized electrons, forming an electronic beam carrying an OAM. The selection rules of photoionization when using a twisted beam carrying an OAM have been theoretically investigated in different configurations in \cite{piconoe2010,matulajpb2013} confirming that, for electrons out of the beam singularity, i.e. far from the focus, the electric dipole transition should remain dominant \cite{suppmat}. 
To test the validity of the electric dipole approximation, we performed an interferometric measurement of such electron beams, building on the well-established XUV-IR cross-correlation technique called RABBIT (reconstruction of attosecond beatings by interferences of two-photon transitions) used to characterize attosecond pulse trains, see e.g. \cite{mairessescience2003}. In RABBIT, a target gas is photoionized by the harmonic comb in the presence of a weak dressing IR field. The simultaneous absorption of an XUV harmonic photon ($\mathrm{H_{q}}$) and one IR photon produces sidebands ($\mathrm{SB_{q+1}}$) in the electron spectrum at energies corresponding to even multiples of the driving laser frequency. The same sidebands are also populated by the absorption of an XUV photon from the following harmonic ($\mathrm{H_{q}}$) and the stimulated emission of one IR photon. The interferences between these two quantum paths lead to oscillations of the sideband yield as a function of the XUV/IR delay (${\tau _0}$) at twice the IR laser period. The phase of these oscillations is directly linked to the total relative phases of the two consecutive harmonics involved and to that of the two-photon dipole transition at play, dubbed as the "atomic phase". Most importantly, for the interferences to appear, the wavefronts of the IR and XUV need to be matched, such as a well-defined phase relation between the XUV and IR field across different transverse parts of the gas jet is obtained. Using a twisted XUV photoionizing beam and assuming that the photoionized electron beam is also twisted (which would be a consequence of the validity of the electric dipole approximation, see \cite{suppmat}), this condition would be met with a dressing IR beam which is also twisted, but not with a regular Gaussian dressing beam. To be more specific, let us assume, as measured above, that the spatial phase of the q-th harmonic is ${\phi _q}(r,\theta ) = {\ell _q}\theta $, where we use the $(r,\theta )$polar coordinates. The measured sideband intensity is an average of the contributions of each emitted electron in the interaction region. If we write the intensity coming from one point in this region, neglecting the contributions of the atomic phases we get \cite{mairessescience2003}:
\begin{eqnarray*}
{\rm{S}}{{\rm{B}}_{q + 1}}\left( {\omega ,\theta } \right) = {\rm{cos}}\left[ {2\omega {\tau _0} + {\varphi _{q + 2}} - {\varphi _q} + \left( {{\ell _{q + 2}} - {\ell _q} - 2{\ell _1}} \right)\theta } \right]
\end{eqnarray*}
where $\omega$ is the angular frequency of the driving laser and ${\varphi _q}$ the spectral phase of the qth order. To keep the intensity oscillating after integration over $\theta$, which is the operation mode of our detector, the $\theta$-dependent term must vanish, giving ${\ell _{q + 2}} - {\ell _q} = 2{\ell _1}$. This condition is only met when taking ${\ell _q} = q{\ell _1}$. \\

We tested this prediction on the experimental setup based on a Mach-Zehnder interferometer schematized in Figure~\ref{fig:two} (A). We could verify that even after reflection off a drilled mirror, the IR focus kept a donut shape and so did the harmonics, with again a constant divergence. Imaging the foci in the sensitive region of the electron spectrometer, we observed a thick donut profile for the dressing beam, ensuring a homogeneous dressing of the harmonics. A spectrum with sidebands lines was then observed (Figure~\ref{fig:three} (A)). Scanning the XUV-IR delay, we observed $2\omega$ oscillations of the sideband intensities. The observation of these oscillations supports i) that the XUV beam carries an OAM and has therefore a ring intensity pattern and ii) that the electric dipole approximation remains valid. As in standard RABBIT measurements, the phase of these oscillations is directly linked to the group delay (GD) of the EWP generated by the harmonics comb. This GD is plotted in Figure~\ref{fig:three} (B).  A linear dependence is obtained, corresponding to group delay between successive harmonics of $\Delta {t_e} = 103 \pm 9$as. This value is almost equal to what has been reported with Gaussian beams (${\ell _1} = 0$) for such a driving peak intensity \cite{mairessescience2003}.

The two sets of measurements reported above, i.e. the spatial intensity profile of the harmonics and their group delay finally provides us with all data needed to reconstruct the full spatio-temporal shape of the emitted attosecond electron beam. We hereafter focus on the case of photoelectron with linear momentum along the propagation axis, copying the spring structure of the XUV. Here we suppose a flat cross section of argon, which is a good approximation in this energy range. As lately predicted, we get two intertwined spring like structures \cite{parienteol2015}. The helical shape is directly related to the ionizing XUV light helicity, while the presence of two such structures is reminiscent of the generation of odd harmonics only. A cut in the spatial domain displays a double lobe structure, while the temporal profile shows a pulse train with each pulse lasting about 200 as. An interesting characteristic of this structure is that it leads to an identical attosecond pulse train when observed at different azimuthal positions, only delayed by half a period of the driving laser (1330 as) in 180 degrees, i.e. 7.4 as/degree. As shown in \cite{hernandezprl2013}, the corkscrew structure still holds in the case of single attosecond pulses. We believe that this property makes such pulses powerful tool for transient absorption spectroscopy experiments, which require tunable pump-probe delays on the attosecond timescale. Here, one may spatially map the attosecond time-delay in a single shot, getting rid of any stability requirement. The dynamical range for the delay scan is here 1.33 fs, i.e. the infrared laser half optical cycle, which could easily be increased up to several femtoseconds by increasing the driving field wavelength.

\section{Conclusions and outlook}
The experimental evidence presented here for the transfer of OAM over an extremely broad spectral range down into the XUV region confirms recent theoretical predictions and ends ongoing controversies about this transfer law. In addition, the analysis method identified here could easily be generalized to arbitrary large spectral bandwidth, opening the study of a variety of even less conventional experimental situations. Based on these grounds, of promising interest is the reported observation of the attosecond structure of the generated XUV pulse train obtained through the measurement of its spectral phase. First, it confirms the conservation of the usual photoionization selection rules with helically phased XUV light beams. While we did not observe any specific behavior for the photoionization of a noble gas, whether or not the presence of OAM might change the photoelectron emission times in more complex systems is unknown and deserves more theoretical investigations. Second, it opens the route to the manipulation of attosecond electron beams carrying OAM’s, which will use the large panel of attosecond tools developed in the past ten years, may it be through high harmonics spectroscopy or XUV-IR pump-probe experiments. In particular, specific dichroisms were predicted in the XUV spectral range and still remain to be observed. Multicolor HHG using such tailored beams could also provide all in one pump-probe schemes, taking advantage of the spatial encoding of the delay in the azimuthal phase with these beams.  Finally, the precise characterization of these new light beams will pave the way to the study of the controversial coupling between SAM and OAM in matter during photoionization.

\bibliography{BiblioLaguerre}
  \cleardoublepage
  \twocolumngrid

\part*{Supplementary Materials} 
\section*{Materials and Methods}
The experimental setup is pictured in Figure~\ref{fig:two} (A). The experiments were performed using the LUCA laser server at CEA Saclay which delivers 30 mJ, 50 fs, 800 nm pulses at 20 Hz. The laser mode was converted using a 16-level spiral phase plate (SPP) manufactured by SILIOS Technologies. \\ For RABBIT measurements, the laser is split in two uneven parts by a mirror with a 8 mm hole. The main (outer) part of the beam is focused by a 1m focal length lens (L) into a gas jet provided by a piezoelectric driven valve (Attotech). A diaphragm (D) removes the remaining IR beam, while the harmonics are focused by a 0.5 m focal length toroidal mirror (TM) into the sensitive region of a 1 m long magnetic bottle electron spectrometer time of flight (MBES). A $\mathrm{SiO_2}$ plate (SP) serves as further attenuation of the IR beam. An extra full weak IR beam (4 mm diameter, energy of 70 $\mathrm{\mu J}$) may be superimposed with the XUV beam in the (MBES) with a delay controlled by a piezoelectric transducer (TS). \\

For intensity measurements, the drilled mirrors are replaced by plain mirrors and L by a 2m focal length lens. The harmonics are collected downstream the MBES on a photon spectrometer made of a variable line spacing Hitachi grating (001-0437) and a micro channel plate (MCP) coupled to a phosphor screen. In order to reveal the 2D spatial profile of the harmonics while spectrally resolving them, the MCP are placed 8 cm before the focal plane of the grating. The phosphor screen is imaged by a Basler A102f CCD camera. The observation distance from the source to the MCP is 115 cm.\\

\section*{S1 - Direct measurement of the OAM of the driving laser at the focal spot}
\begin{figure}
	\renewcommand{\thefigure}{S1}
		\includegraphics[width=0.55\textwidth]{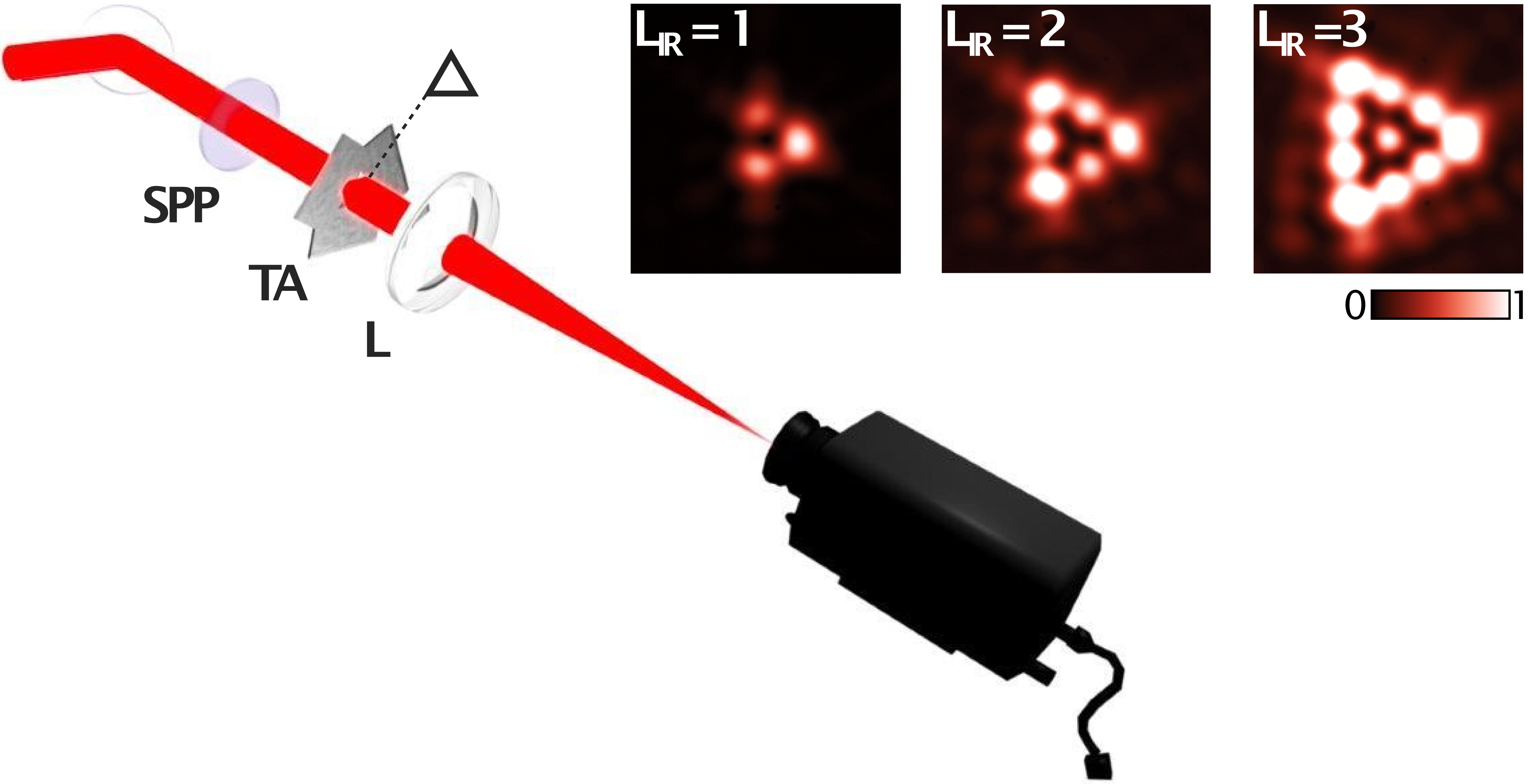}
		\caption{\label{fig:sone}\textbf{Measurement of the OAM of the driving laser. } The collimated and spatially filtered femtosecond laser beam propagates through one or two of the spiral phase plates (SPP). A variable equilateral triangular slit is placed about 10 cm downstream. The diffracted beam is then focused by the same lens as for the HHG experiments. The focus is imaged on a Imagine Source CCD camera equipped with a x10 objective and a 160 mm long tube. The images recorded when placing the masks imposing a $2\pi$, $4\pi$ and $6\pi$ azimuthal phases (corresponding respectively to$\ell  = \hbar $, $\ell  = 2\hbar $ and $\ell  = 3\hbar $) are displayed as insets. In the latter case, two masks imposing $2\pi$ and $4\pi$ are twinned to yield a $6\pi$ phase per turn.}
\end{figure}

The measurement of the OAM in the visible domain usually relies on interferometric or diffractive schemes \cite{yaoaop2011}. We here chose to check the behavior of our phase masks placing a triangular slit in the path of the beam. The experimental scheme used for HHG is just slightly modified: a triangular slit is placed before the focusing lens and the beam at focus is imaged on a CCD camera equipped with a microscope objective. As expected from the theory, when choosing a size of the variable slit comparable to the waist of the laser, we obtained a series of spots along a triangular pattern, whose number is related to the OAM carried by the beam (see Figure~\ref{fig:sone}. This observation was only made when the incoming beam presented a very close to flat wavefront. For instance, constraining a mirror in its mount ruins this structure. This online diagnosis secured the quality of the incoming beam for HHG. 

\section*{S2 - 4D numerical simulation layout}
In this paragraph we provide details about the simulation of HHG using a driving laser carrying an OAM. For the IR beam, we accounted for the full experimental setup described below and schematized in Figure~\ref{fig:two}. The structure of the LG mode, which does not possess the cylindrical symmetry of a Gaussian beam, required performing 4D (3D in space+time) simulations. To this end, the beam is propagated through the different optical elements, by means of the Huygens-Fresnel integral, up to the gas jet entrance. This is the starting point for HHG calculations. In brief the coupled propagation equations for the driving laser and harmonic fields are numerically solved on a three dimensional spatial grid, in the paraxial and slowly varying envelope approximations, using a standard finite-difference method. Calculations are performed in a frame moving at the group velocity of the laser pulse, i.e. the speed of light in vacuum here. Both electron and atomic dispersions are taken into account. We first compute the driving field at a given time t in the pulse envelope and position z along the propagation axis. This field is then used for calculating the space- and time-dependent ionization yields and dipole strengths. Ionization rates are modeled using Ammosov-Delone-Kraïnov (ADK) tunneling formula \cite{ammosovjetp1986} while dipoles are computed in the SFA, following the model described in \cite{lewensteinpra1994}. Once these two terms are obtained, the equation of propagation for the harmonic field is solved. The calculation is repeated for each t and z. The spiral phase plate inducing the OAM is modeled by $\varphi (x,y) = {\ell _1}{\tan ^{ - 1}}\frac{y}{x}$, where x and y are the coordinates in the plane perpendicular to the propagation axis.

\section*{S3 - Analytical expression of the ring diameter for an OAM beam}
Using cylindrical coordinates $(r,\theta ,z)$, where r is the distance from the optical axis, z the distance along this axis from the focal point and ${z_R} = \pi w_0^2/{\lambda _1}$ the Rayleigh range, ${w_0}$ the beam waist and $\lambda_1$ the wavelength of the radiation, the spatial profile of the electric field of a LG mode reads, up to normalization constants 
                                   \begin{align*}
																	{E_\ell }\left( {r,\theta ,z} \right) =& {\rm{\;}}\frac{{{{\rm{e}}^{{\rm{i}}\ell \theta }}}}{{w\left( {\lambda ,\;z} \right)}}{\left( {\frac{r}{{w\left( {\lambda ,\;z} \right)}}} \right)^{\left| \ell  \right|}}\\&\times{e^{\left( { - {\rm{i}}\frac{{{\rm{\pi }}{r^2}z}}{{{\lambda _1}\left( {{z_R}^2 + {z^2}} \right)}} - \frac{{{r^2}}}{{w{{\left( {\lambda ,\;\xi } \right)}^2}}} - i\psi _{\ell ,0}^L} \right)}},  \tag{S1}
																	\label{eq:sone} 
																	\end{align*}
where $\ell $ corresponds to the OAM per photon in this mode, $w(\lambda ,z) = {w_0}\sqrt{1+{(z/{z_R})}^2}$ is the equivalent of the waist for a Gaussian beam, and $\psi _{\ell ,0}^L = \left( {\left| \ell  \right| + 1} \right){\tan ^{ - 1}}(z/{z_R})$is the equivalent of the Gouy phase. Let us consider HHG driven by such an infrared LG beam and assume that the q-th order is also a LG mode with an OAM ${\ell _q}$. The corresponding intensity writes

	\[
	{I_\ell }(r,\theta ,z) = \frac{1}{{{w^2}\left( {\lambda,z} \right)}}{\left( {\frac{r}{{w\left( z \right)}}} \right)^{2\left| \ell  \right|}}{e^{\left( { - \frac{{2{r^2}}}{{{w^2}\left( {\lambda ,z} \right)}}} \right)}}
		\]
The maximum intensity along the radius is obtained at $\partial {I_\ell }/\partial r = 0$, i.e. 
	\[
	\left( {\frac{{2\left| \ell  \right|}}{r}{\rm{\;}} - \frac{{4r}}{{{w^2}\left( z \right)}}} \right){\rm{\;}}{\left( {\frac{r}{{w\left( z \right)}}} \right)^{2\left| \ell  \right|}}{e^{\left( { - \frac{{2{r^2}}}{{{w^2}\left( z \right)}}} \right)}} = 0
	\]	
The solution to this equation is: 
	${\rm{\;}}{r_{{\rm{max}}}} = w\left( {\lambda ,z} \right)\sqrt {\frac{{\left| \ell  \right|}}{2}} $	
For each order, the maximum intensity is located at:
\begin{align*}
{r_{{\rm{max}}}}\left( {{\lambda _q},z} \right) =& \;w\left( {{\lambda _q},z} \right)\sqrt {\left| {\frac{{{\ell _q}}}{2}} \right|} \\ =& \;w\left( {{\lambda _q},0} \right)\sqrt {1 + {{\left( {\frac{{\pi zw{{\left( {{\lambda _q},0} \right)}^2}}}{{{\lambda _q}}}} \right)}^2}} \sqrt {\left| {\frac{{{\ell _q}}}{2}} \right|} 
\end{align*}
where ${\lambda _q} = {\lambda _1}/q$ is the wavelength of the q-th order. We suppose that LG IR driven HHG occurs at focus of a lens (z=0). Without giving any detail about the generation process, it can be assumed that the intensity profile of the XUV beam is maximal where the intensity of the IR field is maximal well. This is predicted in the Strong Field Approximation (SFA) \cite{lewensteinpra1994} or directly by solving the Time Dependent Schrödinger Equation (TDSE) \cite{kulander1992}. Equating the two maxima of the IR and the XUV yields \[w\left( {{\lambda _q},0} \right)\sqrt {\frac{{{\ell _q}}}{2}}  = w\left( {{\lambda _1},0} \right)\sqrt {\frac{{{\ell _1}}}{2}}. \] Taking into account$\;{\lambda _q} = {\lambda _1}/q$, one gets: 
\[
	{r_{{\rm{max}}}}\left( {q,z} \right) = w\left( {{\lambda _1},0} \right)\sqrt {\left| {\frac{{{\ell _1}}}{2}} \right|} .\sqrt {1 + {{\left( {\frac{{{\rm{\pi }}zw{{\left( {{\lambda _1},0} \right)}^2}}}{{{\lambda _1}}}\frac{{q{\ell _1}}}{{{\ell _q}}}} \right)}^2}} ,
	\]
which is the expression given in the text. We retrieve the fact that the OAM of the beam rules its divergence. This is also obtained when using the generalized Hankel transform to calculate the propagation of the beam as detailed in the following section. 

\section*{S4 - Propagation of an OAM beam using the generalized Hankel transform}
In this section we describe how the propagation of XUV light carrying OAM can be computed efficiently. With this treatment in hand, we can numerically compute harmonic profiles with all possible values of OAM carried by each harmonic. We then find which solution for the “transfer law” matches our experimental data, which shows a constant ring diameter in the far-field over an arbitrary spectral range.\\
 
The propagation of the harmonic radiation to the far field can be described using the standard Huygens-Fresnel integral. The electric field in a plane located at a distance z along the axis reads:
\begin{widetext}
	\[{E_{z,\;{\lambda _q}}}\left( {x,y} \right) = \frac{{{e^{\;\;i\frac{{\rm{\pi }}}{{{\lambda _q}z}}\left( {{x^2} + {y^2}} \right)}}}}{{i{\lambda _q}z}}\int \limits_{ - \infty }^\infty \int \limits_{ - \infty }^\infty  {E_0}\left( {{x_0},{y_0}} \right)\;\;{{\rm{e}}^{\;i\frac{\pi }{{{\lambda _q}z}}\left( {{x_0}^2 + {y_0}^2} \right)}}{{\rm{e}}^{\;\frac{{ - i2{\rm{\pi }}\left( {x{x_0} + y{y_0}} \right)}}{{z{\lambda _q}}}}}{\rm{d}}{x_0}{\rm{d}}{y_0},\]
\end{widetext}	
\noindent with $\left( {{x_0},{y_0}} \right)$ and $\left( {x,y} \right)$ being the Cartesian coordinates respectively  at focus and at the z position, ${\lambda _q}$ the wavelength of the considered harmonic, and ${E_0}$ the harmonic electric field at focus. In the case of light carrying OAM, this integral can be advantageously rewritten by considering the symmetry of the problem. First we change to the cylindrical coordinates $\left( {R,\phi } \right)$ for the far-field and $\left( {r,\theta } \right)$ at focus:
\begin{widetext}
	\[{E_{z,{\lambda _q}}}\left( {R,\phi } \right) = \frac{{{e^{\;i\frac{{\rm{\pi }}}{{{\lambda _q}z}}{R^2}}}}}{{i{\lambda _q}z}}\int \limits_0^\infty \int \limits_0^{2{\rm{\pi }}} {E_0}\left( {r,\theta } \right){e^{i\frac{{\rm{\pi }}}{{{\lambda _q}z}}{r^2}}}{e^{\;\frac{{ - i2{\rm{\pi }}Rr\cos \left( {\theta  - \phi } \right)}}{{z{\lambda _q}}}}}r{\rm{d}}r{\rm{d}}\theta \]
	\end{widetext}
	
\begin{figure}
\renewcommand{\thefigure}{S2}
		\includegraphics[trim=0 2cm 0 2cm,clip=true,width=0.55\textwidth]{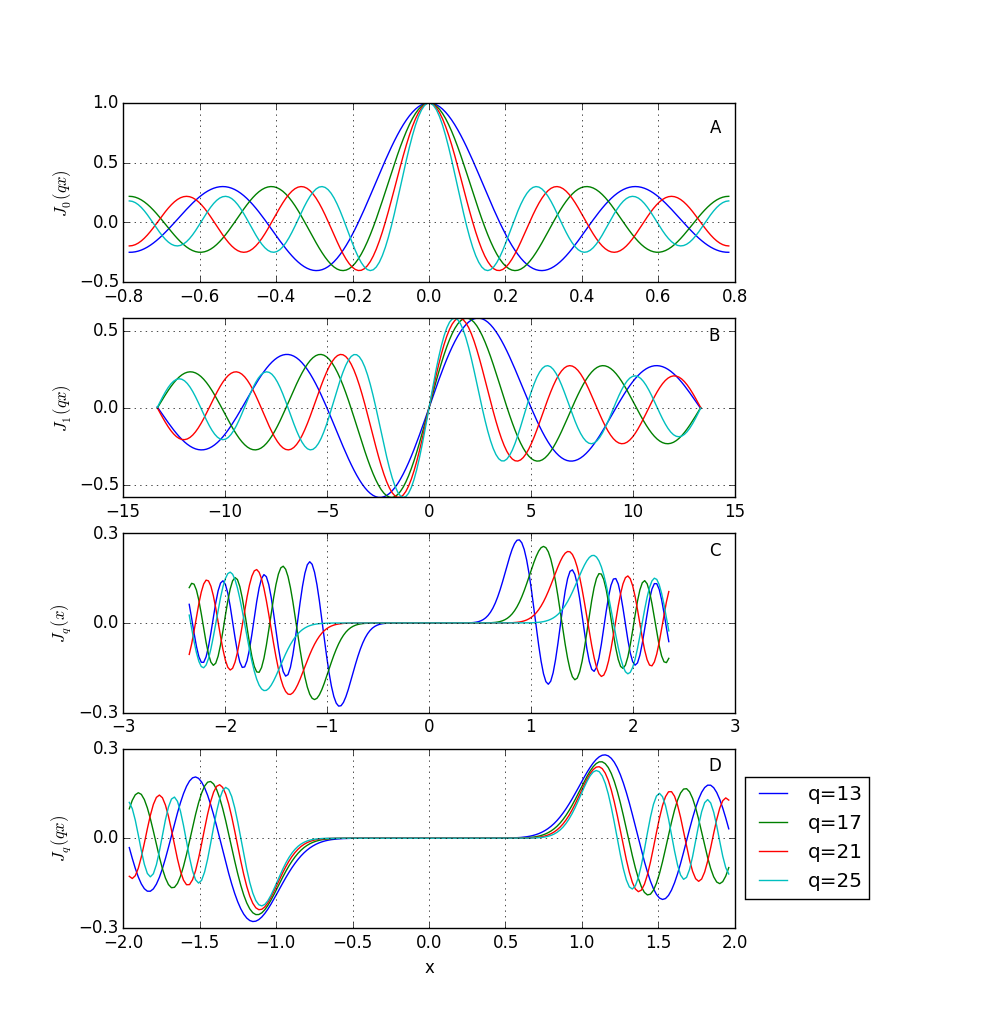}
		\caption{\label{fig:stwo}Computation of the zeroth (A), first (B) and qth order (C-D) Bessel functions for arguments qx (A, B and D) and x (C). The x-scales have been adjusted to yield a few oscillations in each panel. The orders are q=13, 17, 21 and 25.}
\end{figure}

In the case of HHG with a driving field carrying OAM, symmetries may be invoked to write the XUV field at focus as ${E_0}\left( {r,\theta } \right) = \;{U_0}\left( r \right){e^{i{\ell _{\rm{q}}}{\rm{\theta }}}}$, where ${U_0}\left( r \right)$ is the r-dependent part of the complex field at focus. This yields:
\begin{align*}
	{E_{z,{\lambda _q}}}\left( {R,\phi } \right) =& \frac{{{e^{i\frac{{\rm{\pi }}}{{{\lambda _q}z}}{R^2}}}}}{{i{\lambda _q}z}}\int \limits_0^\infty  {U_0}\left( r \right){e^{i\frac{{\rm{\pi }}}{{{\lambda _q}z}}{r^2}}} \\
	&\int \limits_0^{2{\rm{\pi }}} {e^{i{\ell _{\rm{q}}}{\rm{\theta }}}}{e^{\frac{{ - i2{\rm{\pi }}Rr\cos \left( {\theta  - \phi } \right)}}{{z{\lambda _q}}}}}r{\rm{d}}r{\rm{d}}\theta 
\end{align*}
Using the identity${\rm{\;\;}}{{\rm{J}}_n}\left( x \right) = \frac{1}{{2{\rm{\pi }}{i^n}}}\mathop \smallint \limits_0^{2{\rm{\pi }}} {e^{ix\cos \alpha }}{e^{in\alpha }}{\rm{d}}\alpha $, where ${{\rm{J}}_n}$ is the n-th order Bessel function of the first kind \cite{wolf1979}, we can carry out the integration over ${\rm{\theta }}$ and obtain the intensity in the far field:
		\begin{align*}
		\label{eq:stwo}
		{I_{z,{\lambda _q}}}\left( {R,\phi } \right) = {\left| {\frac{{2{\rm{\pi }}}}{{{\lambda _q}z}}\mathop \smallint \limits_0^\infty  {U_0}\left( r \right){e^{i\frac{{\rm{\pi }}}{{{\lambda _q}z}}{r^2}}}{{\rm{J}}_{{\ell _{\rm{q}}}}}\left( {\frac{{2{\rm{\pi }}Rr}}{{z{\lambda _q}}}} \right)r{\rm{d}}r} \right|^2}
			\tag{S2}
			\end{align*}
This integral now takes the well-known form of a Hankel transform. The ${{\rm{J}}_n}\left( x \right)$ functions show aperiodic oscillations whose zeros location increases with ${\ell _{\rm{q}}}$. In the following we track the origin of the constant divergence observed for beams carrying an OAM down to the overlap of ${U_0}\left( r \right)$ and  ${{\rm{J}}_{{\ell _{\rm{q}}}}}\left( {\frac{{2{\rm{\pi }}Rr}}{{z{\lambda _q}}}} \right).$ A few relevant Bessel functions are displayed in Fig.~\ref{fig:stwo}.\\

\begin{figure*}
\renewcommand{\thefigure}{S3}
		\includegraphics[width=0.9\textwidth]{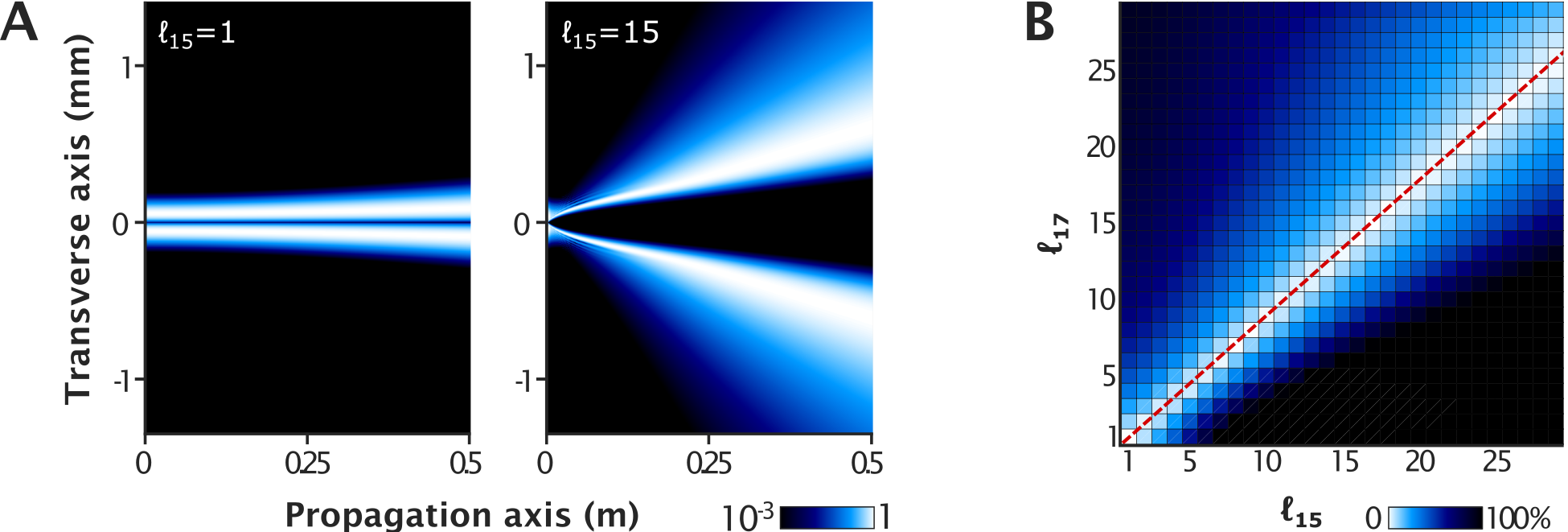}
		\caption{\label{fig:sthree}\textbf{Computation of the Hankel transform in Eq.~\ref{eq:stwo}.} The waist of the harmonic field at focus is taken to be 60 $\rm{mu m}$. \textbf{(A)} Intensity of the 15-th harmonic during propagation along the optical axis. The field is propagated to the far-field using equation (S2) from z = 0 to z = 0.5 m. The role of OAM in the field divergence is illustrated by taking either ${\ell _{15}} = 1$ (left) or ${\ell _{15}} = 15$ (right). The intensity is normalized to one for every value of z and is plotted in logarithmic scale.\\
\textbf{(B)} Relative difference between the diameters of H15 and H17 as a function of their respective OAM. The far-field profile of H15 and H17 are computed using (S2) and their diameter are gathered for every value of ${\ell _{15}}$ and ${\ell _{17}}$ between 1 and 30. For every couple of (${\ell _{15}},{\ell _{17}}$) values, the difference of their diameter is plotted in percentage of the diameter of H15. The dashed red curve is a linear fit of the combinations for which this difference is minimal. It corresponds to ${\ell _{17}} = 1.13{\ell _{15}} \simeq \frac{{17}}{{15}}{\ell _{15}}$. The same was done for all harmonics from q=13 to q=25, which yielded the general result ${\ell _{q + 2} = \frac{{q + 2}}{q}\ell_q}$.}
\end{figure*}

\noindent For ${\ell _q}$= 0, which is the usual case of HHG, the variation of the extension of ${I_{z,{\lambda _q}}}\left( {R,\phi } \right)$ with $\lambda q$ is ruled both by the scaling of ${{\rm{J}}_0}\left( {\frac{{2{\rm{\pi }}Rr}}{{z{\lambda _q}}}} \right)$ with q and the phase terms in ${U_0}\left( r \right)$. ${U_0}\left( r \right)$ has a distribution of amplitude centered about r=0 in this case. Since${\rm{\;\;}}{{\rm{J}}_0}\left( {\frac{{2{\rm{\pi }}Rr}}{{z{\lambda _q}}}} \right) = 1$ at R = 0 (Fig.~\ref{fig:stwo} (A)), the main contribution to the integral thus come from this central part of ${U_0}\left( r \right).$ However${\rm{\;\;}}{{\rm{J}}_0}\left( {\frac{{2{\rm{\pi }}Rr}}{{z{\lambda _q}}}} \right)$ varies little with the harmonic order (Fig.~\ref{fig:stwo} (A)) and thus plays a minor role in the scaling with q. On the contrary, ${U_0}\left( r \right)$ shows a large spherical phase term, whose curvature significantly varies with q. This is reminiscent of i) the driving laser intensity profile at focus, which is parabolic as a first approximation and ii) the dependence of the excursion of the electron in the continuum with the amplitude of the electric field, which has a first order term. Considering this phase, it is found that the beam diameter increases with q for the shortest electronic trajectories. 

\noindent Now, when considering ${\ell _q} \ne 0$ and a ring intensity distribution for $\;{U_0}\left( r \right)$, the result is completely different. Like in the case of  ${\ell _q} = 0$, within the thickness of the ring, the phase of ${U_0}\left( r \right)$ is spherical. However, now the Bessel functions are zero at R = 0 and the main contributions to the integral come from outer parts of the intensity distribution.  If ${\ell _q} = {\ell _1}$ , the variation of the Bessel function with q remains significant (Fig.~\ref{fig:stwo} (B)). One thus expects the ring diameter to vary with q. On the contrary, with ${\ell _q} = q \times {\ell _1}$, and an argument proportional to q (Fig.~\ref{fig:stwo} (D)), the position and width of the first lobe is almost stationary with q. This is the origin of the constant ring diameter observed. Note that this is the result of an interplay between the scaling factor of the argument and the order of the Bessel function (compare Fig.~\ref{fig:stwo} (C) and (D)).\\

To put this analysis on firmer grounds, we used Eq.~ref{eq:stwo} to propagate a given field at focus with a chosen OAM value by carrying out a pth-order quasi-discrete Hankel transform \cite{guizarjosa2004}. We used the standard radial distribution for a LG mode as $\rm{U_0(r)}$. The result of the computation varying z is reported in Fig.~\ref{fig:sthree} (A) for the 15-th harmonic when choosing either ${\ell _{15}} = 1$ or ${\ell _{15}} = 15$. The effect of the OAM carried by the beam is dramatic, suggesting that the measurement of the diameter of the rings is highly sensitive to this OAM.
We repeated this computation systematically for harmonic orders 13 to 25 and for ${\ell _q}$ = 1 to 30 for each harmonic, finally getting a 2D matrix of ring diameters indexed by harmonic numbers and OAM values.  The relative variations of these values were compared to each other. An example of such a comparison for orders 15 and 17 is displayed in Fig.~\ref{fig:sthree} (B). Clearly, a line of minimal variation appears, which nicely fits the slope$\;\;{\ell _{17}} \simeq \frac{{17}}{{15}}{\ell _{15}}$ . This was observed for all other couples of harmonic orders (q, q+2): it appears that the ring diameter stays the same only when$\;\;{\ell _{q + 2}} = \frac{{q + 2}}{q}{\ell _q}$, which yields for any harmonic order$\;\;{\ell _q} = q{\ell _1}$. This demonstrates that the multiplicative law for OAM transfer through HHG is the only one matching our experimental observations.
	
\section*{S5 - Spiral wavefronts and attosecond light springs}
In this paragraph we intend to give physical insights on the multiplicative law of OAM established in the main text and on the formation of light springs with a double spiral structure. \\

\begin{figure}
\renewcommand{\thefigure}{S4}
		\includegraphics[width=0.5\textwidth]{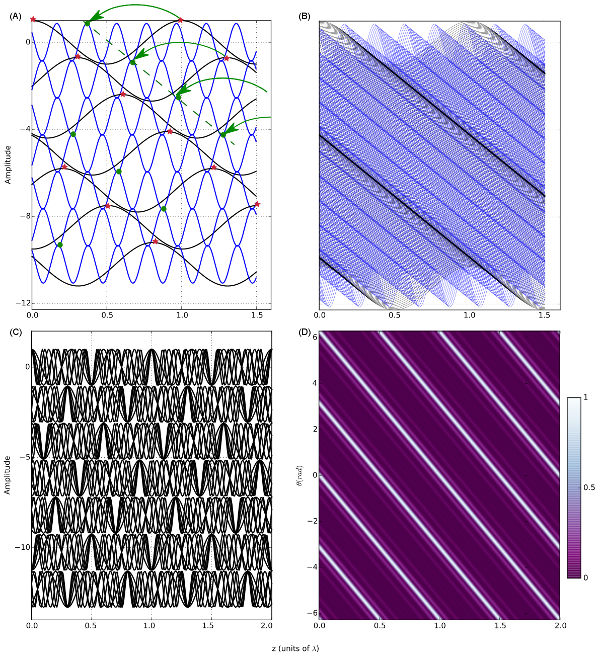}
		\caption{\label{fig:sfour}\textbf{(A-B)} Plot of a series of progressively offset $A{\rm{cos}}\left( {\frac{{2{\rm{\pi }}}}{\lambda }z - \ell \theta  + \varphi } \right)$ curves against z at different azimuths $\theta$ going from 0 to $2\pi$ from top to bottom. Black curves: $\lambda_1$, A=1, $\ell $=1 and $\phi=\pi/3$. Blue curves  $5\lambda_1$, A=0.9, $\ell $=5 and $\phi=\pi/3$. On the left (resp right) panel, 7 (resp. 200) angles from $\theta=0$ to $\theta=2\pi$ are represented. The red stars on the left panel are positioned at the maxima of the field, where tunnel ionization may occur, and green circles are where recombination and XUV emission occurs. The right panel shows 5 blue caustics, corresponding to the 5 intertwined spiral phase fronts. \textbf{(C)} Same as panel (A) but for 5 harmonics ($3, 5\lambda_1, 7\lambda_1, 9\lambda_1, 11\lambda_1$)) for each angle. \textbf{(D)} Color map of the intensity of the field resulting of the superposition of these five harmonics against the propagation axis (horizontal) and the azimuthal angle (vertical). }
\end{figure}

The origin of the multiplicative law may be inferred analyzing the HHG process in the time domain. We consider the simplest microscopic model of HHG, by which an electron first tunnels out of an atom close to an extremum of the driving field, second acquires energy by quivering in this field and finally recombines with its parent ion emitting its excess of energy as XUV radiation \cite{corkum2007}. The phase slip of the emission with respect to the driving laser extremum (denoted ${\varphi _{{\rm{XUV}}}}$) is firstly determined by the last two steps of the process, namely the time spent by the electron in the continuum and the phase of the recombination dipole. We denote T the laser period and temporarily consider an infinitely thin plane of radiating dipoles perpendicular to the main propagation axis of the laser beam. At different azimuths $\theta$, due to the structure of the wavefront of the beam carrying an OAM, the extrema of the field are shifted in time by ${\ell _1}\theta T/2\pi $ (red stars in Figure~\ref{fig:sfour} (A)). The emission time of the XUV radiation (recombination time of the electron) at azimuth $\theta$ will thus write ${t_e}(\theta ) = ({\varphi _{XUV}} + {\ell _1}\theta )T/2\pi $, showing a single spiral shape (green dots in Figure~\ref{fig:sfour} (A)). In space, at a given time, the pitch of this spiral is $\lambda_1$. To have an OAM $q \times {\ell _1}$, we should have $q \times {\ell _1}$such intertwined spirals within  $\lambda_1$. To retrieve this observation we need to consider i) that the medium is long compared to the wavelength and ii) that we are using a multi-cycle driving field (about 40 extrema considering a temporal FWHM of 50 fs, as used in this work). If we consider emissions at both maxima and minima of all half cycles of the driving field when it propagates, the same sketch could be drawn T/q later. This delay is chosen so that at a given azimuth $\theta$ the two emissions at time 0 and T/q later add up coherently. At T/q, the black curves representing the driving field, together with the red stars and green circles attached to them appear dephased by T/q while the blue curves remain unchanged. This leads to a second dashed green line parallel to the first, offset by $\lambda_1/q$. As apparent on the right panel (Fig.~\ref{fig:sfour} (B)), the final harmonic field thus shows a series of q intertwined spiral phase fronts, corresponding, as expected, to an OAM $q \times {\ell _1}$. 

Finally, coherently superposing a series of successive odd harmonics having such a multiple spiral phase shape results in the destructive interference of most of these spirals, only two of them per cycle remaining (Fig S4 C and D). This is the origin of the intertwined double spiral shape displayed in Fig.~\ref{fig:three} of the main text.

\section*{S7 -	Experimental spectra using neon as a generating gas}
\begin{figure}[!h]
\renewcommand{\thefigure}{S5}
		\includegraphics[width=0.5\textwidth]{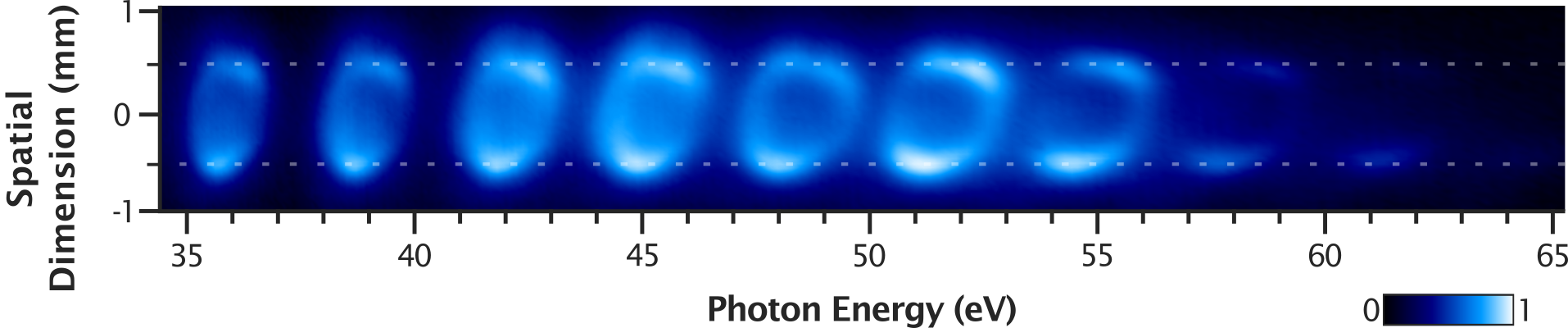}
		\caption{\label{fig:ssix}Normalized intensity of harmonics $\mathrm{23^{rd}}$ to $\mathrm{41^{st}}$ generated in neon and observed in the far field, using $\ell=1$. The dashed white lines represent the average ring diameter.}
\end{figure}

High harmonics were generated in a neon gas target using ${\ell _1} = 1$ for the generating beam. As seen in Fig.~\ref{fig:ssix}, we were able to generate harmonics up to 65 eV, corresponding to the 41-st order. The intensity exhibits a very nice ring profile, and we measure the average diameter to be $1.00 \pm 0.05$ mm, which is the same as what was measured in argon. This confirms our former observations on a broader spectral range. The harmonics presenting a constant divergence throughout the spectrum, the multiplicative law for OAM transfer is verified. The highest harmonic therefore carries an OAM of 41, while having a regular intensity profile. 

\section*{S8	- Rabbit with OAM’s }

\begin{figure*}
\renewcommand{\thefigure}{S6}
		\includegraphics[width=0.7\textwidth]{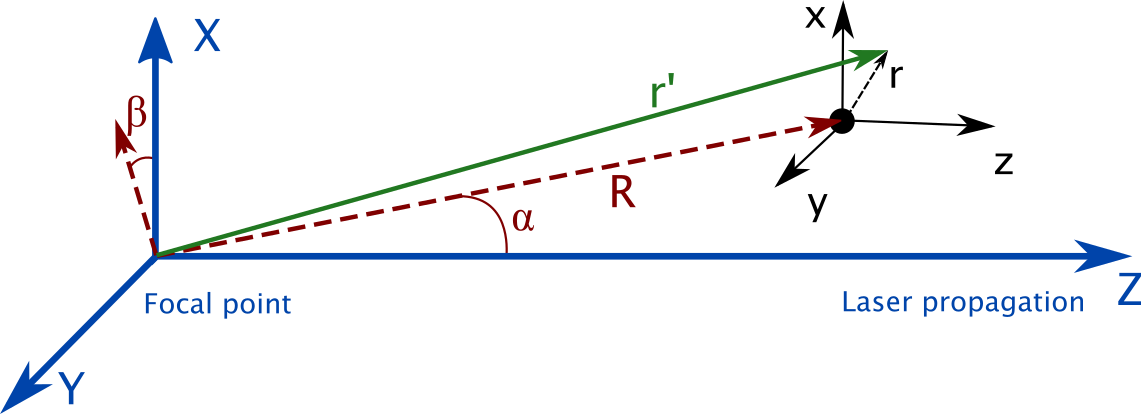}
		\caption{\label{fig:sseven}Geometry of the ionization of an atom by a twisted light beam. The coordinates (X,Y,Z) refer to the laboratory frame and (x,y,z) to the atomic frame centered on the atom nucleus.}
\end{figure*}

Here we propose to analyze how the OAM of a twisted photon beam transfers to electrons through photoemission of atoms and investigate the validity of the dipolar selection rule. The formalism and notations below are closely related to the ones used in \cite{piconoe2010}.\\

\noindent We first define the vectors which represent the electron and nucleus positions in the laboratory and the atomic frames, respectively. In the laboratory frame, the nucleus is located at $\vec{\rm {R}} = \left( {{\rm{R}},{\rm{\alpha }},{\rm{\beta }}} \right)$ and the electron at $\vec {\rm{r'}} = \left( {{\rm{r'}},{\rm{\theta '}},\phi {\rm{'}}} \right)$, while in the atomic frame the nucleus is at (0, 0, 0)  and the electron at ${\vec{\rm{r}}} = \left( {{\rm{r}},{\rm{\theta }},\phi } \right)$. The two sets of coordinates are related by ${\vec {\rm{r'}}} = \vec{\rm {R}} + {\vec{\rm{r}}} .$ Rewriting Eq.(2) of \cite{piconoe2010} for $p = 0$ and using the vectors defined above, we obtain for the expression for the vector potential at the electron position ${\vec {\rm{r'}}}$:
\begin{widetext}
\begin{align}
\label{eq:sthree}
\vec A\left( {\vec r',t} \right){\rm{\;}} \propto \frac{1}{{w\left( {z'} \right)}}{\left( {\frac{{\rho '}}{{w\left( {z'} \right)}}} \right)^{{\rm{|}}{\ell _1}|}}exp\left( { - \frac{{{{\rho '}^2}}}{{{w^2}\left( {z'} \right)}}} \right)exp\left[ {i{\ell _1}\phi ' - i\frac{{{\rho ^2}z'}}{{{w^2}\left( {z'} \right){z_R}}} - i\psi _{\ell ,0}^L\left( {z'} \right)} \right].\; \tag{S3}
\end{align}	
\end{widetext}
Here, we have ${\rm{\rho '}} = {\rm{r'sin\theta '}}$,  ${\rm{z'}} = {\rm{r'cos\theta '}}$,  and the other quantities are the same as in Eq.~\ref{eq:sone}.	\\ \\

The photoelectron angular distribution for the absorption of one photon is given by the differential cross section which, according to the Fermi Golden rule, requires computing such a matrix element
\begin{align}
\label{eq:sfour}
\int {{\rm{d}}^3} {\vec{\rm{r}}} \; {\rm{\psi_f^*}} {(\vec{\rm{r}})} {\left[{\vec{\rm{A}}} \left({\vec{\rm{r'}}},\rm{t}\right).{\vec{\rm{p}}}\right]} {\rm{\psi_i^*}} {(\vec{\rm{r}})},
\tag{S4}
\end{align}


\noindent where ${{\rm{\psi }}_{\rm{i}}}\left({\vec{\rm{r}}} \right), {{\rm{\psi }}_{\rm{f}}}\left( {\vec{\rm{r}}} \right)$ are the initial and final state wavefunctions, respectively. Two limiting cases appear: (i) ${\rm{Rsin\alpha }} = 0$; (ii) ${\rm{Rsin\alpha }} >  > {\rm{r}}$ ($\approx 1 \rm{ a.u}$). In the first one, we recover the study of \cite{piconoe2010} with $\phi {\rm{'}} = \phi ,{\rm{\;}}$which means that the angles appearing in Eq.~\ref{eq:sthree} are the same as the ones in the integral of Eq.~\ref{eq:sfour}. Therefore, one obtains nondipolar selection rules (Eq. (6) of \cite{piconoe2010}) which depend on the value ${\ell _1}$ of the OAM, reminiscent of the marked variations of A at the atomic scale, in the vicinity of the beam axis. In the other case, we can assume $\vec {{\rm{r'}}} \approx {\vec{\rm{R}}}$, and retrieve the dipole approximation. There, ${\vec{\rm{A}}}\left( {\vec{\rm{r'}}},{\rm{t}} \right)$ at the atomic scale becomes independent of the electron’s coordinates and just acts as a position-dependent amplitude and phase factor, ${\rm{\xi }}\left( {\vec{\rm{R}}} \right)$ on the usual 
dipole matrix element. It leads to the standard $\Delta L = \pm1$ rule, which is also used in different experiments involving single twisted photon transferring its OAM to an ensemble of atoms \cite{yaoaop2011}.
This implies that, apart from a tight region in space close to the laser propagation axis, where the electric field is very small, twisted photons act like "normal" (i.e. plane-wave) ones, except that they carry a ${\vec{\rm{R}}}$-dependent phase, which adds up in multiphoton processes like HHG or RABBIT. Thus, the q-th harmonic will have an OAM of ${\ell _{\rm{q}}} = {\rm{q}} \times {\ell _1}$ and the ${\vec{\rm{R}}}$-dependent RABBIT oscillations of the sideband will be modulated by
\begin{eqnarray*}
{\rm{S}}{{\rm{B}}_{{\rm{q}} + 1}}\left( {{\rm{\omega }},{\vec{\rm{R}}}} \right) = {\rm{cos}}\left[2{\rm{\omega }}{{\rm{\tau }}_0} + {{\rm{\varphi }}_{{\rm{q}} + 2}} - {{\rm{\varphi }}_{\rm{q}}} \right. \\
\left. + \left( {{\ell _{{\rm{q}} + 2}} - {\ell _{\rm{q}}} - 2{\ell _{\rm{d}}}} \right){\rm{\xi}}\left({\vec{\rm{R}}}\right)\right],
\end{eqnarray*}

\noindent where ${\ell _d}$ is the OAM of the dressing field and ${\varphi _{q + 2}},{\rm{\;}}{\varphi _q}$ the usual harmonic phase. If the dressing beam is homogeneous and shows almost flat phase fronts (${\ell _d} = 0$ for instance) the last phase term covers $0-2\pi$ and washes out the oscillations when averaging over the positions of the atom, as a temporal jitter would do. On the contrary, dressing with an IR carrying an OAM ${\ell _d} = \frac{{{\ell _{q + 2}} - {\ell _q}}}{2}$ ensures keeping the oscillatory behavior of the sidebands.

\end{document}